\documentclass[aps,showpacs,floats,twocolumn,floatfix,superscriptaddress]{revtex4-1}
 \pdfoutput=1
\setcitestyle{super}
\pdfoutput=1
\usepackage{mathrsfs}
\usepackage{bm,amsbsy,amssymb,amsmath}
\usepackage{caption}
\usepackage{subcaption}
\usepackage{graphics,graphicx,dcolumn,fleqn,epic,eepic,float,tabularx}
\usepackage{multirow,rotate,rotating,color}
\usepackage[utf8]{inputenc}
\newcommand{\figref}[1]{Fig.~\ref{fig:#1}}
\newcommand{\eqnref}[1]{Eq.~(\ref{eq:#1})} 
  \definecolor{tuered}{RGB}{214,0,74}
  \definecolor{tueblue}{RGB}{0,102,204}
  
\newcommand{\revisedtext}[1]{\textcolor{black}{#1}}
\usepackage{tikz}
\usepackage{relsize}
\tikzset{fontscale/.style = {font=\relsize{#1}}}
\usetikzlibrary{calc}

\begin{document}
\title{Controlled Capillary Assembly of Magnetic Janus Particles \\
 at Fluid-Fluid Interfaces}
 \author{Qingguang Xie}
  \email{q.xie1@tue.nl}
  \affiliation{Department of Applied Physics, Eindhoven University of Technology, P.O. Box 513, NL-5600MB Eindhoven, The Netherlands}
  \author{Gary B. Davies}
  \email{gbd@icp.uni-stuttgart.de}
  \affiliation{Institute for Computational Physics, University of Stuttgart,\\ Allmandring 3, 70569 Stuttgart, Germany}
  \author{Jens Harting}
  \email{j.harting@tue.nl}
  \affiliation{Forschungszentrum J\"ulich, Helmholtz Institute Erlangen-N\"urnberg for Renewable Energy (IEK-11), F\"urther Straße 248, 90429 N\"urnberg, Germany}
  \affiliation{Department of Applied Physics, Eindhoven University of Technology, P.O. Box 513, NL-5600MB Eindhoven, The Netherlands}
\date{\today}

\begin{abstract}

Capillary interactions can be used to direct assembly of particles adsorbed at fluid-fluid interfaces.
Precisely controlling the magnitude and direction of capillary interactions to assemble particles into favoured structures for materials science purposes is desirable but challenging.
In this paper, we investigate capillary interactions between magnetic Janus particles adsorbed at fluid-fluid interfaces.
We develop a pair-interaction model that predicts that these particles should arrange into a side--side configuration, and carry out simulations that confirm the predictions of our model.
Finally, we investigate the monolayer structures that form when many magnetic Janus particles adsorb at the interface.
We find that the particles arrange into long, straight chains exhibiting little curvature, in contrast with capillary interactions between ellipsoidal particles.
We further find a regime in which highly ordered, lattice-like monolayer structures form, which can be tuned dynamically using an external magnetic field.   

\end{abstract}

\maketitle

\section{Introduction}
Colloidal particles absorb strongly at fluid-fluid interfaces because the attachment energy is much greater than the thermal energy.~\cite{Binks2001}
Particle properties such as weight,~\cite{Chan1981} roughness,~\cite{Adams2008} and shape anisotropy~\cite{Loudet2005} can deform the interface around the adsorbed particle. Recent studies have shown that external electric~\cite{Aubry2008b} or magnetic fields~\cite{Gary2014a, Gary2014b} can also cause particles to deform the interface.
When interface deformations of individual particles overlap, the fluid-fluid surface area varies, leading to capillary interactions between the particles.~\cite{Madivala2009,Park2011,Lewandowski2010,Aubry2008a}

The shape of the interface deformations around individual particles characterises the capillary interaction modes that occur between the particles. 
The modes can be represented analytically 
as different terms in a multipolar expansion of the interface height around the particle:~\cite{Stamou2000}
particle weight triggers the monopolar mode~\cite{Kralchevsky1994}, external torques switch on the dipolar mode,~\cite{Gary2014a} and particle surface roughness and shape anisotropy activate the quadrupolar mode.~\cite{Botto2012}

The arrangement of many particles adsorbed at a fluid-fluid interface depends on both the dominant capillary interaction mode and how strongly an individual particle deforms the interface. The dipolar and quadrupolar modes are anisotropic, which allows the possibility of directed assembly. Varying both the dominant mode and/or the magnitude of the interface deformations can profoundly change the assembly behaviour of particles at interfaces. For example, Yunker et al.~\cite{yunker_suppression_2011} showed that switching spherical particles to ellipsoidal particles induces quadrupolar capillary interactions between the particles that inhibit the coffee-ring effect. However, these interactions depend only on particle properties and are therefore difficult to control on-the-fly assembly.

A desirable next step is to control the capillary strength or dominant capillary mode dynamically, allowing far greater control of the particle assembly process. Recently, Vandewalle et al.~\cite{Vandewalle2013} used heavy spherical particles with a magnetic dipole to tune the interplay between capillary attraction and magnetic repulsion. Applying an alternating magnetic field even causes the particles to swim across the interface, creating so-called magneto-capillary swimmers. 

Recently, Davies et al.~\cite{Gary2014a,Gary2014b} showed that the dipolar mode is a promising route to
dynamically-controlled anisotropic assembly at an interface. They \revisedtext{simulated} ellipsoidal particles with a dipole moment parallel to the particle's long-axis and applied an external magnetic field normal to the interface. The dipole-field interaction causes the particles to tilt with respect to the interface, which deforms the interface, and the degree of deformation can be controlled by the dipole-field strength. They further demonstrated that one can switch off the capillary interactions altogether by causing the particles to flip into a vertical orientation with respect to the interface in which no interface deformations exist. 

It is highly desirable to create ordered chains or crystals of particles at interfaces for materials science purposes. The structures created using both the dipolar mode and the quadrupolar mode tend to form open-chains with little long-range order.~\cite{Loudet2005, Gary2014b} Xie et al.~\cite{Xie2015} recently showed how to create tunable dipolar capillary interactions using spherical magnetic Janus particles adsorbed at fluid-fluid interfaces. 

In this paper, we simulate the interaction between many such Janus-capillary particles at an interface, and find that we can create highly-ordered chains of particles. Additionally, we develop a theoretical model based on the superposition approximation describing capillary interactions between two particles, and measure the capillary forces between them, finding the results in good agreement with our theoretical model. 

This paper is organised as follows: Section~\ref{sec:methods} briefly describes our simulation methods before we present our results in Section~\ref{sec:results}, and Section~\ref{sec:conclusions} concludes the article.

\section{Simulation method}
\label{sec:methods}
We use the lattice Boltzmann method (LBM) to simulate the motion of each fluid. 
The LBM is a local mesoscopic algorithm, allowing for efficient parallel implementations, and has demonstrated itself as a powerful tool for numerical simulations of fluid flows.~\cite{Succi2001}
It has been extended to allow the simulation of, for example, multiphase/multicomponent 
fluids~\cite{Shan1993,Shan1994,LKLSNJWVH16} and suspensions of particles of arbitrary shape and wettability.~\cite{ladd-verberg2001, Jansen2011, Gunther2013a}

We implement the pseudopotential multicomponent LBM method of Shan and Chen~\cite{Shan1993} with a D3Q19 lattice~\cite{Qian1992} and review some relevant details in the following.
Two fluid components are modelled by the following evolution equation of each distribution function discretized in space and time according to the lattice Boltzmann equation:
\begin{eqnarray}
  \label{eq:LBG}
 f_i^c(\vec{x} + \vec{c}_i \Delta t , t + \Delta t)=f_i^c(\vec{x},t)+\Omega_i^c(\vec{x},t)
  \mbox{,}
\end{eqnarray}
where $i=1,...,19$, $f_i^c(\vec{x},t)$ are the single-particle distribution functions for fluid component $c=1$ or $2$, 
$\vec{c}_i$ is the discrete velocity in the $i$th direction, and 
\begin{equation}
  \label{eq:BGK_collision_operator}
  \Omega_i^c(\vec{x},t) = -\frac{f_i^c(\vec{x},t)- f_i^\mathrm{eq}(\rho^c(\vec{x},t), \vec{u}^c(\vec{x},t))}{\left( \tau^c / \Delta t \right)}
\end{equation}
is the Bhatnagar-Gross-Krook (BGK) collision operator.~\cite{Bhatnagar1954} $\tau^c$ is the relaxation time for component $c$. 
The macroscopic densities and velocities are defined as 
$ \rho^c(\vec{x},t) = \rho_0 \sum_if^c_i(\vec{x},t)$, where $\rho_0$ is a reference density, and $\vec{u}^c(\vec{x},t) = \sum_i  f^c_i(\vec{x},t) \vec{c}_i/\rho^c(\vec{x},t)$, respectively.
Here, $f_i^\mathrm{eq}(\rho^c(\vec{x},t),\vec{u}^c(\vec{x},t))$ is a third-order equilibrium distribution function.~\cite{Chen1992}
When sufficient lattice symmetry is guaranteed, the Navier-Stokes equations can be recovered from \eqnref{LBG} on appropriate length and time scales~\cite{Succi2001}.
For convenience we choose the lattice constant $\Delta x$, the timestep $ \Delta t$, the unit mass $\rho_0 $ and the relaxation time $\tau^c$ to be unity, which leads 
to a kinematic viscosity $\nu^c$ $=$ $\frac{1}{6}$ in lattice units.

For fluids to interact, we introduce a mean-field interaction force between fluid components $c$ and $c'$ following the Shan-Chen approach.~\cite{Shan1993}
The Shan-Chen LB method is a diffuse interface method, resulting in an interface width of $\approx 5\Delta x$.
The particle is discretized on the fluid lattice and coupled to the fluid species by means of 
a modified bounce-back boundary condition as pioneered by Ladd and Aidun.~\cite{ladd-verberg2001, AIDUN1998, HFRRWL14}
For a detailed description of the method including the extension to particles suspended in multicompoent flows, 
we refer the reader to the relevant literature.~\cite{Jansen2011, Gunther2013a, Frijters2012, Cappelli2015, KFGKH13}

We perform simulations of two particles and multiple particles.
For simulations between two particles, we place two particles along the $x$ axis separated by a distance $L_{AB}$. 
We fix the position and orientation of the particles and let the system equilibrate. 
We then measure the lateral forces on the particles as a function of tilt angle and bond angle, respectively.
For simulations of multiple particles, we randomly distribute particles at the fluid-fluid interface, 
and let the system equilibrate. \revisedtext{We note that there is no thermal coupling with the fluid and 
no fluctuating hydrodynamics involved in our simulation method.} 
We then apply a magnetic field and analyse the results after the system reaches a steady-state.

\section{Results and Discussion}
\label{sec:results}
\subsection{Pair interactions}
\label{sec:pair}
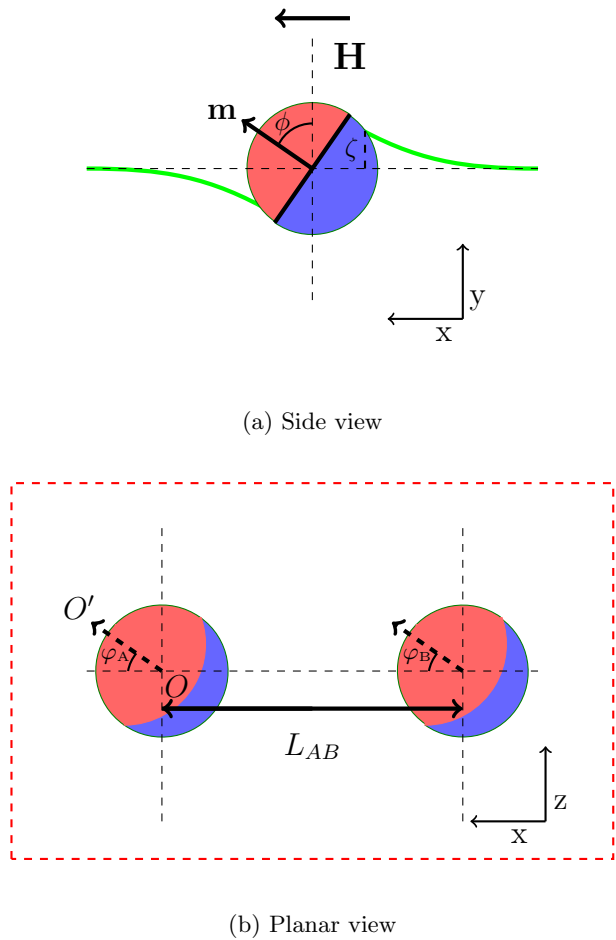
\begin{figure}[!h]
\begin{subfigure}{.5\textwidth}
 \begin{tikzpicture}

	\fill[white,opacity=0.4] (-3,-3) rectangle (3.0,0.0);
	\fill[white,opacity=0.4] (-3,0) rectangle (3.0,3.0);
	
	\draw[ultra thick,green] (0.5,0.6) to [out=330,in=180] (3,0.0);
	\draw[ultra thick,green] (-0.5,-0.6)  to  [out=150,in=360](-3,0.0);

	 \filldraw[fill=blue!60!white, draw=green!50!black]

      (0,0) -- (-0.5,-0.72) arc (235:415:0.87) -- (0,0);
    \filldraw[fill=red!60!white, draw=green!50!black]
      (0,0) -- (0.5,0.72) arc (55:235:0.87) -- (0,0);
	
	\draw[dashed] (-3,0) to (3,0);
	
	\draw[ultra thick, black,arrows=->] (0.5,2.0) to (-0.5,2.0);
	\node at (0.5,1.5) [fontscale=3] {$\mathbf{H}$};

    	\draw[thick,black] ([shift=(0:0.5cm)]-0.5,0.6) arc (90:155:0.5cm);
  	\node[black] at (-0.4,0.6) [fontscale=1] {$\phi$};
  	\draw[ultra thick, black,arrows=->] (0,0) to (-0.94,0.65);
  	\node at (-1.2,0.75) [fontscale=2] {$\mathbf{m}$};

  	\draw[ultra thick, black] (-0.5,-0.72) -- (0.5,0.72); 
  	\draw[thick, black, dashed] (0.7,0.0) -- (0.7,0.5);  

  	\node at (0.52,0.25) [fontscale=1] {$\zeta$};
  	\draw[dashed] (0,-1.75) to (0,1.75);

\draw[thick, black,arrows=->] (2,-2) to (2,-1.0);
\node at (2.2,-1.75) [fontscale=2] {y};
 \draw[thick, black,arrows=->] (2,-2) to (1.0,-2);
  \node at (1.75,-2.2) [fontscale=2] {x};

\end{tikzpicture}

\caption{Side view}
\label{fig:geo}
\end{subfigure}


\begin{subfigure}{.5\textwidth}
 \begin{tikzpicture}

	\fill[white,opacity=0.4] (-4,-3) rectangle (4.0,0.0);
	\fill[white,opacity=0.4] (-4,0) rectangle (4.0,3.0);

       \filldraw[fill=blue!60!white, draw=green!50!black]
      (-2.0,0) -- (-2.5,-0.72) arc (235:415:0.8766) -- (-2.0,0);
    \filldraw[fill=red!60!white, draw=green!50!black]
      (-2,0) -- (-1.5,0.72) arc (55:235:0.8766) -- (-2.0,0);
       \filldraw[fill=red!60!white, draw=red!60!white]
       (-2.6222,0.617465) -- (-2.5,-0.72) arc (-90:19:1.08) -- (-2.6222,0.617465);

	\filldraw[fill=blue!60!white, draw=green!50!black]
      (2.0,0) -- (1.5,-0.72) arc (235:415:0.87) -- (2.0,0);
    \filldraw[fill=red!60!white, draw=green!50!black]
      (2,0) -- (2.5,0.72) arc (55:235:0.87) -- (2.0,0);
       \filldraw[fill=red!60!white, draw=red!60!white]
       (1.3778,0.617465) -- (1.5,-0.72) arc (-90:19:1.08) -- (1.3778,0.617465);
       
	\draw[dashed] (-3,0) to (3,0);
	\draw[red,thick,dashed] (-4,-2.5) rectangle (4,2.5);

	\draw[ultra thick, dashed, black,arrows=->]  (-2.0,0) to (-2.94,0.65);
  	\draw[ultra thick, dashed, black,arrows=->] (2,0) to (1.06,0.65);
  	  \node at (-3.1,0.85) [fontscale=2] {$O'$};
  	   \node at (-1.8,-0.2) [fontscale=2] {$O$};

   	\draw[ultra thick, black,arrows=->] (-2.0,-0.5) -- (2.0,-0.5);
   	\draw[ultra thick, black,arrows=->] (0.0,-0.5) -- (-2.0,-0.5);
   	\draw[dashed] (-2.0,-2) to (-2.0,2);
   	\draw[dashed] (2,-2) to (2,2);
	\node at (-2.61,0.2) [fontscale=0.4] {$\varphi_{\text{\textscale{.5}{A}}}$};
	\draw[thick,black] (-2.45,0) arc (170:140:0.5cm);
	
	\node at (1.4,0.2) [fontscale=0.4] {$\varphi_{\text{\textscale{.5}{B}}}$};
	\draw[thick,black] (1.55,0) arc (170:140:0.5cm);
  \node at (0.0,-1.0) [fontscale=2] {$L_{AB}$};

\draw[thick, black,arrows=->] (3.1,-2) to (3.1,-1.0);
\node at (3.3,-1.75) [fontscale=2] {z};
 \draw[thick, black,arrows=->] (3.1,-2) to (2.1,-2);
  \node at (2.75,-2.2) [fontscale=2] {x};
  
\end{tikzpicture}

\caption{Planar view}
\label{fig:geo}
\end{subfigure}

\caption{ a) Side view of a single particle and b) planar view of both particles. 
The Janus particle consists of an apolar and a polar hemisphere. The particle's magnetic
dipole moment $\mathbf{m}$ is orthogonal to the Janus boundary, and the
external magnetic field, $\mathbf{H}$, is directed parallel to the interface.
The tilt angle $\phi$ is defined as the angle between the particle's dipole moment and the undeformed interface normal.
The bold green line represents the deformed interface and $\zeta$ is the maximal interface height at the contact line.
The bond angle $\varphi=\varphi_A=\varphi_B$ is defined as the angle between the projection of orientation of magnetic dipole 
on the undeformed interface (arrow $OO'$) and center-to-center vector of particles. $L_{AB}$ is pair separation.
}
\label{fig:janus-geo}
\end{figure}

In order to gain insight into the behaviour of large numbers of magnetic Janus particles adsorbed at an interface interacting via capillary interactions, we first consider the interaction of two spherical Janus particles. Each particle comprises an apolar and a polar hemisphere of opposite wettability, with three-phase contact
angles $\theta_a = 90^{\circ} +\beta$  and $\theta_p = 90^{\circ} -\beta$,
respectively, where $\beta$ represents the amphiphilicity of the particle. Each particle has a magnetic moment $\mathbf{m}$ directed perpendicular to the Janus boundary, as illustrated in~\figref{janus-geo}a.

When a magnetic field directed parallel to the interface $\mathbf{H}$ is applied the particles experience a torque $\boldsymbol{\tau}=\mathbf{m} \times \mathbf{H}$ that causes them to rotate. The surface tension of the interface resists this rotation, and the particles tilt with respect to the interface. The tilt angle $\phi$ is defined as the angle between the particle dipole-moment $\mathbf{m}$ and the undeformed-interface normal. 

As the particle tilts, it deforms the interface around it in a dipolar fashion:~\cite{Xie2015} the interface is depressed on one side and elevated on the other, and the magnitude of these deformations are equal. The maximal deformation height of the interface, $\zeta$, occurs at the surface of the particle. 

The tilt angle therefore depends on the dipole-field strength $B_m=|\mathbf{m}| |\mathbf{H}|$. For the purposes of our investigations, we assume that the external field strength is much greater than the magnitude of the dipole-moment, $H \gg m$, such that the external field strength is the dominant contribution the dipole-field strength $B_m \approx H$ and we therefore neglect any magnetic dipole-dipole interactions between the particles. 

Capillary interactions arise when the interface deformations caused by the tilting of one particle overlap with the interface deformations caused by the tilting of another particle. Like deformations attract, and unlike deformations repel. 

To investigate the capillary interactions between two particles quantitatively, we define two bond angles $\varphi_A,~\varphi_B$ as the angle between the projection of orientation of the magnetic dipole on the undeformed interface and the centre-centre vector of particles, as shown in~\figref{janus-geo}b. In this part of the paper, we study the interaction between two Janus particles with equal bond angles $\varphi=\varphi_A=\varphi_B$.

\begin{figure}[!t]
  \includegraphics[width=0.45\textwidth]{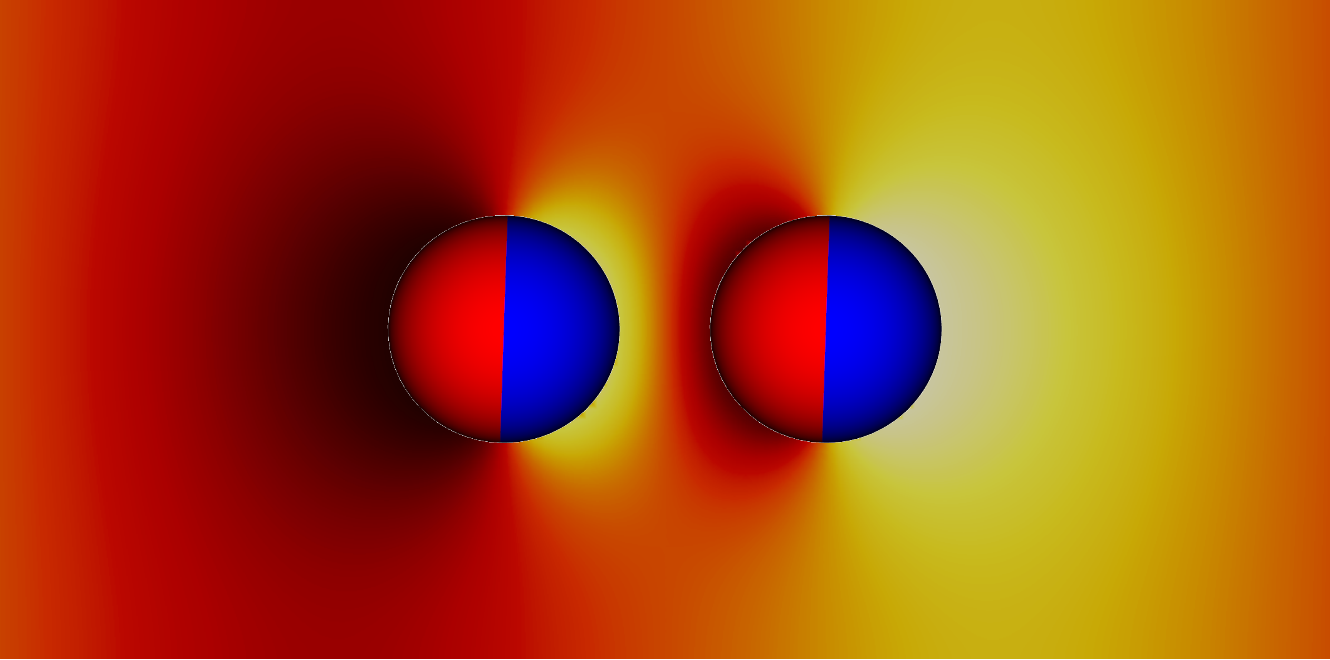} 
 \caption{Snapshots of two tilted Janus particles adsorbed at a fluid-fluid
interface obtained from our simulations. The particles have amphiphilicities
$\beta=39^{\circ}$, tilt angles $\phi = 90^{\circ}$, bond angles $\varphi_A =
\varphi_B = 0^{\circ}$, and a centre-centre separation $L_{AB}/R = 3$. The
colours show the relative height of the interface. The interface is depressed
on one side of the particle (black), raised on the other side (yellow) and flat
at a point equidistant between the two particles. The interface deformations of
each particle are dipolar, causing dipolar capillary interactions between the
particles.}
\captionsetup[figure]{slc=off}
\label{fig:janus-geo-sim}
 \end{figure}

\figref{janus-geo-sim} shows how the interface deforms around two tilting Janus
particles. In this representative system, the particles have equal tilt angles
$\phi=90^{\circ}$ and bond angles $\varphi = 0^{\circ}$.  The
yellow colour represents elevated regions, and the black colour represents
depressed regions of the interface. In this configuration, the particles repel
each other due to the unlike arrangement of their capillary charges. At a point
equidistant between the two particles, the interface deformation is zero.

We have derived an expression for the interaction energy between two Janus
particles interacting via capillary interactions of the kind described
above. In this model, we assume that (i) the leading order deformation
mode is dipolar (ii) the superposition approximation is valid (iii)
interface deformations are small.~\cite{Stamou2000} The dipolar
interaction energy for two Janus particles $\Delta E$ using cylindrical coordinates is
 \begin{align}
  &\Delta E= 2 \pi\zeta^2\gamma_{12}  R^2 L_{AB}^{-2} \nonumber \\
  &+ \frac{8\zeta R^2 \gamma_{12} \sin \beta} {L_{AB}} \left(\tan^{-1}\frac{L_{AB}+R}{L_{AB}-R}-\frac{\pi}{4}\right) \mbox{,} \label{eq:energy}
 \end{align}
where $\gamma_{12}$ is the fluid-fluid interface tension, $R$ is the particle
radius (both particles have the same radii), and $L_{AB}$ is the centre-centre
separation of the particles. To reiterate, $\beta$ is the amphiphilicity of the
particles and $\zeta$ is the height of the maximal interface deformation caused
by the particles. 
The lateral capillary force $\Delta F = \frac{\partial (\Delta E)}{\partial
L_{AB}}$ is therefore
\begin{align}
&\Delta F = -4 \pi \zeta^2 \gamma_{12} R^2 L_{AB}^{-3} \nonumber \\ 
&- \frac{8\zeta R^2 \gamma_{12} \sin\beta} {L_{AB}^2} \left( \frac{L_{AB}R}{R^2+L_{AB}^2} + \tan^{-1}\frac{L_{AB}+R}{L_{AB}-R} - \frac{\pi}{4}\right) \mbox{.}  \label{eq:force-upup-n} 
\end{align}
For a detailed derivation, we refer the reader to Appendix A.

 \begin{figure}
\includegraphics[width= 0.5\textwidth]{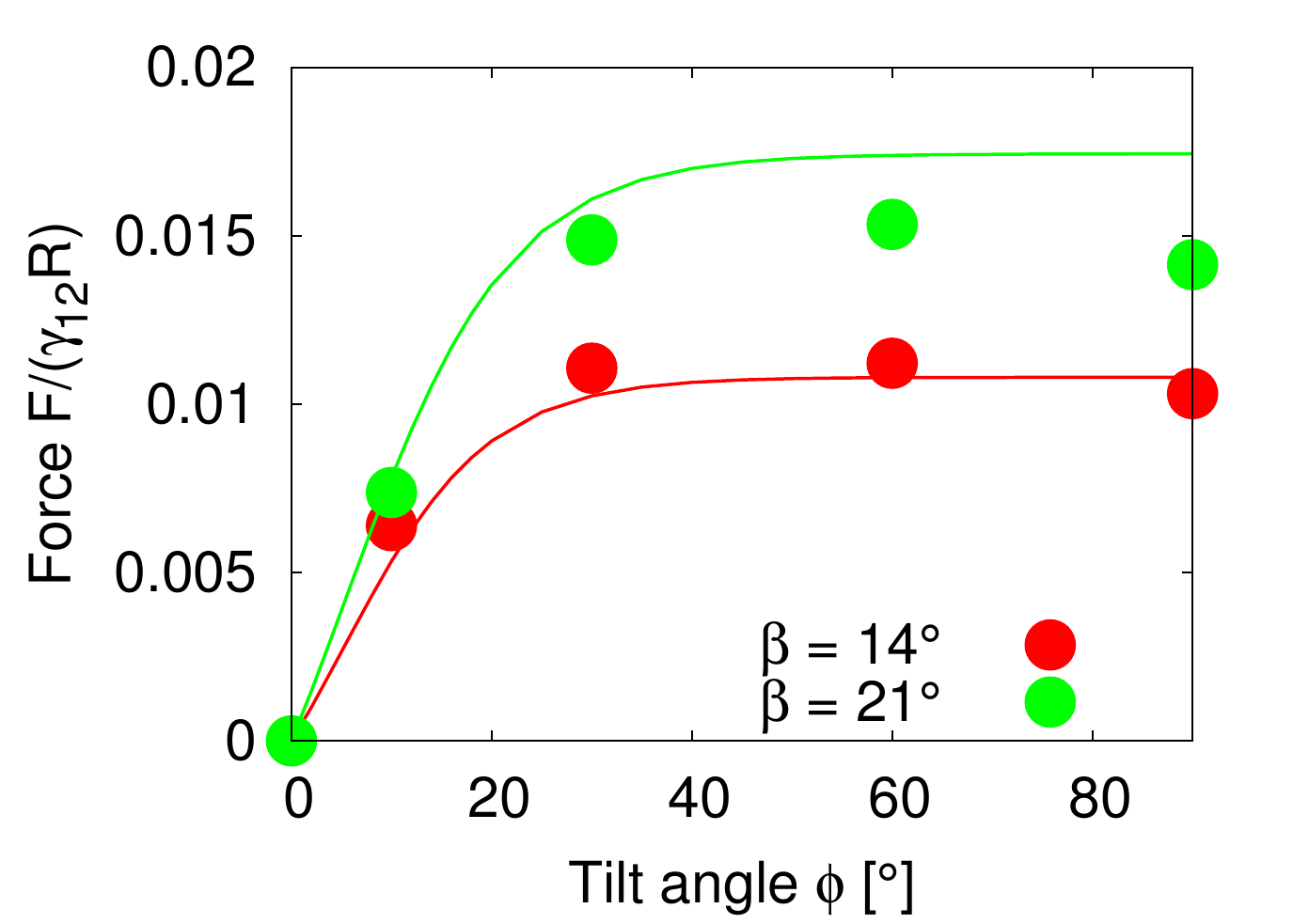}
\caption{Lateral capillary force as a function of tilt angle for particles with amphiphilicities $\beta = 14^{\circ}$ (red) and $\beta = 21^{\circ}$ (green).
The particles have equal bond angles $\varphi_{A}=\varphi_{B}=0$. The solid line represents values from our theoretical model (\eqnref{force-upup-n}), and the symbols are simulation data. 
The theoretical analysis agrees well with our simulation results in the limit of small interface deformations.
}
\label{fig:f-tilt}
\end{figure}

The maximal interface height $\zeta$ in~\eqnref{force-upup-n} depends on the tilt angle $\phi$.
In the case of a single Janus particle, the height of the contact line increases linearly for
small tilt angles, and then reaches a constant value for large tilt angles,
as we reported in our previous work.~\cite{Xie2015} 
\revisedtext{Interestingly, a hyperbolic tangent $\zeta = \tanh(\phi)$ approximates the variation of the contact line height $\zeta$ with tilt angle $\phi$ well.~\cite{Xie2015} }
Therefore, \eqnref{force-upup-n} can be written as a function of the tilt angle. 

Assuming that the interface height $\zeta(\phi) \propto \tanh \phi$,~\cite{Xie2015} 
We compare the theoretical lateral capillary force \eqnref{force-upup-n} (solid lines) to the measured lateral capillary force from our simulations (circles)  in~\figref{f-tilt} for two different particle amphiphilicities $\beta = 14^{\circ}$ and $\beta = 21^{\circ}$. 

We place two particles of radius $R=14$ a distance $L_{AB} = 60$ apart along the $x$-axis with total system size $S = 1536 \times 384 \times 512$. We fix the bond angle $\varphi=0^{\circ}$ between the particles and measure the lateral force on the particles as the tilt angle varies.

~\figref{f-tilt} shows that the lateral capillary force increases as the amphiphilicity increases from $\beta=14^{\circ}$ to $\beta=21^{\circ}$ for a given tilt angle. For a given amphiphilicity, the capillary force increases with tilt angle up to tilt angles $\phi \approx 30^{\circ}$. This is because the interface area increases for small tilt angles,~\cite{Xie2015} which increases the interaction energy. As the tilt angle increases further $\phi > 30^{\circ}$, the capillary force tends to a nearly constant value, due to the fact that the maximal contact line height (and therefore the deformed interface area) also tends to a constant value.~\cite{Xie2015} 

When comparing our theoretical model (solid lines) with simulation data (circles), we see that our model captures the qualitative features of the capillary interaction well, and quantitatively agrees with the numerical results for small tilt angles $\phi<25^{\circ}$ and small amphiphilicities $\beta=14^\circ$. The quantitative deviations at large tilt angles in the $\beta=21^\circ$ case are due to the breakdown of various assumptions in the theoretical model, namely the assumption of small interface slopes, and of finite-size effects in our simulations. The important predictions of our model are that the capillary force between particles can be tuned by increasing the particle amphiphilicity and/or the particle tilt angle. Since the external field strength controls the tilt angle, this allows the tuning of capillary interactions using an external field.

\begin{figure}
\includegraphics[width= 0.5\textwidth]{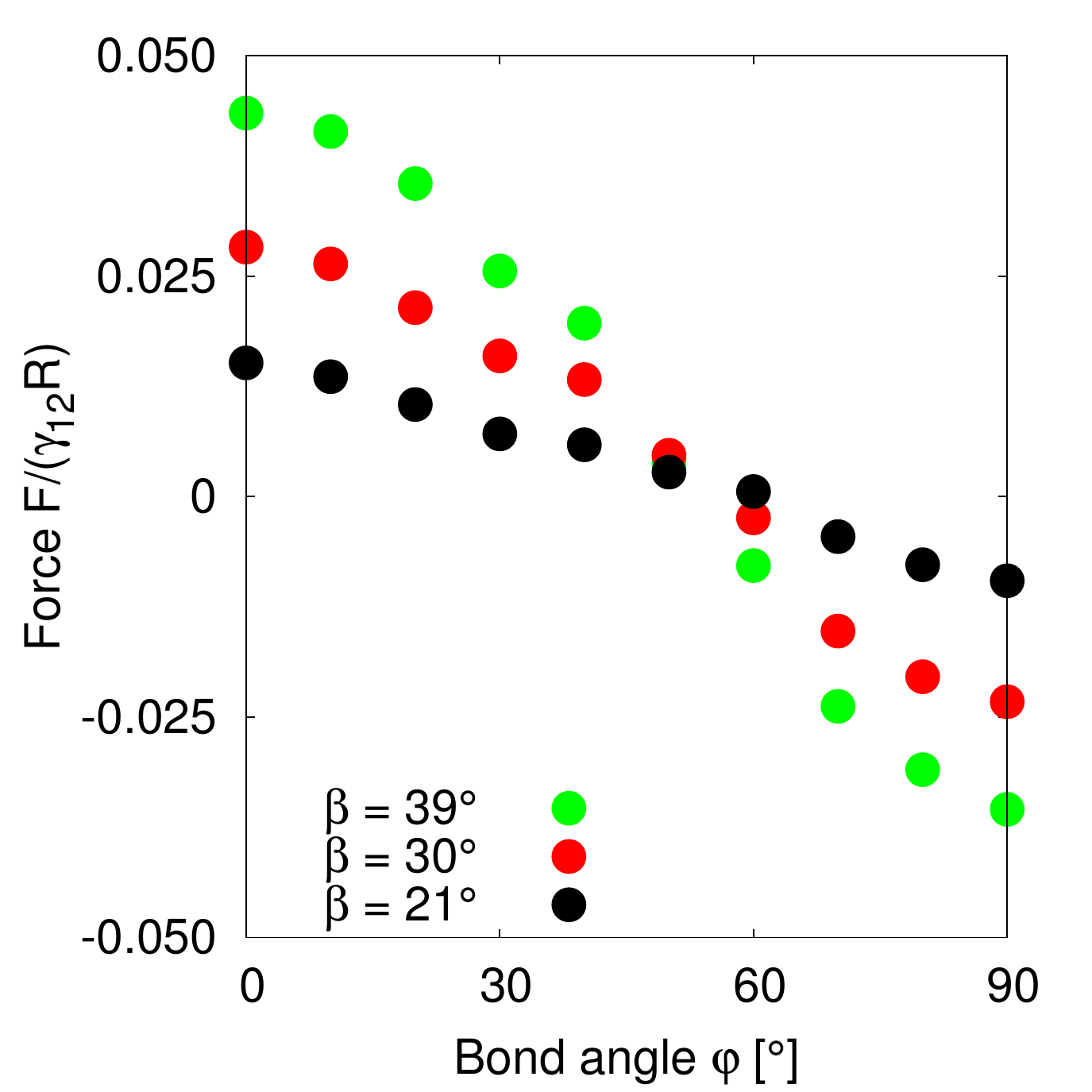}
\caption{Lateral capillary force as a function of bond angles of particles 
with amphiphilicity $\beta = 21^{\circ}$(red), $\beta = 30^{\circ}$(green) and $\beta = 39^{\circ}$(blue).
The particles have tilt angle $90^{\circ}$. The capillary force is repulsive for small bond angles, 
and becomes attractive for large bond angles. There is maximal attractive force between two particles for bond angles $\varphi=90^{\circ}$.
}
\label{fig:f-bond}
\end{figure}
\begin{figure*} [bht]
\begin{center}
\begin{tabular}{m{0.20cm}m{3.0cm}m{3.0cm}m{3.0cm}m{3.0cm}m{3.0cm}}
 &\includegraphics[width= 0.18\textwidth]{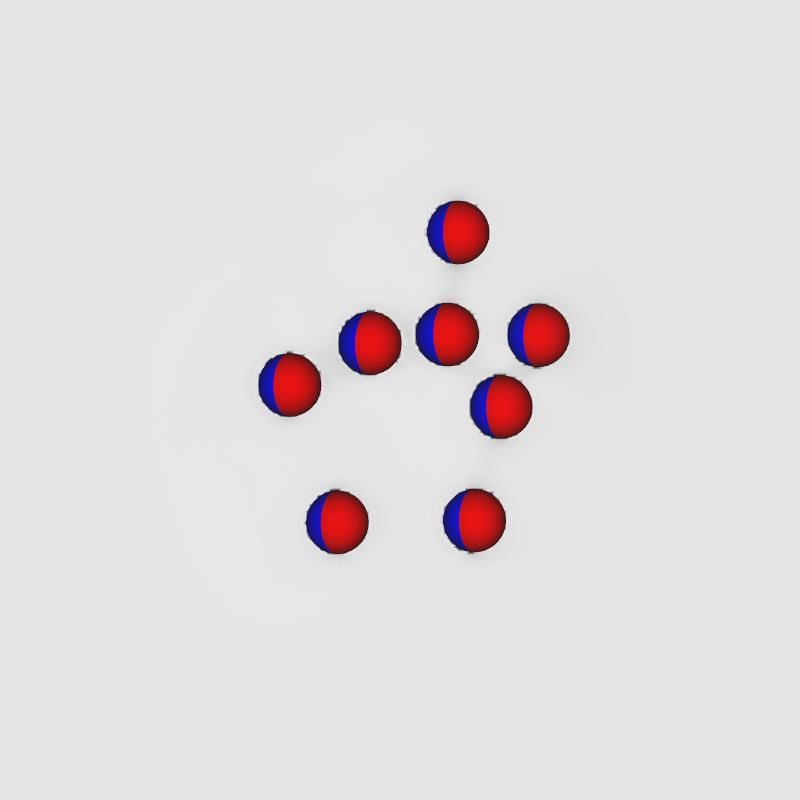}
 \subcaption{}
 \label{fig:n8-0} 
 &\includegraphics[width= 0.18\textwidth]{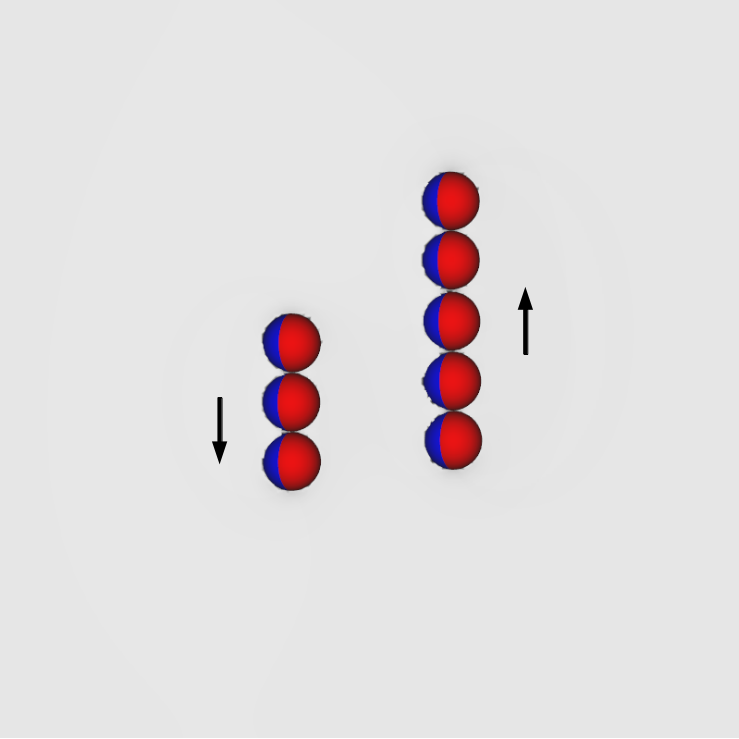}
 \subcaption{}
 \label{fig:n8-1} 
  &\includegraphics[width= 0.18\textwidth]{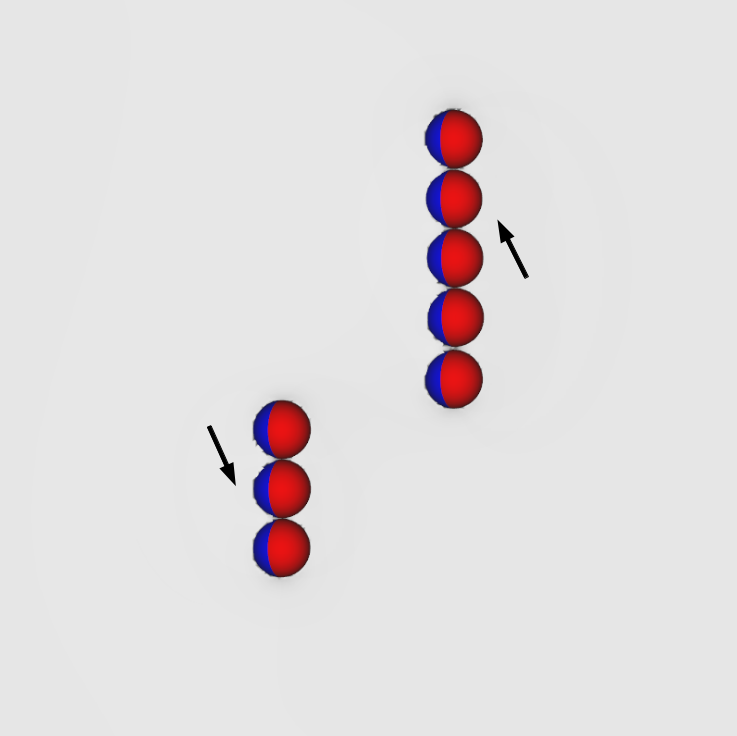}
 \subcaption{}
 \label{fig:n8-2} 
 &\includegraphics[width= 0.18\textwidth]{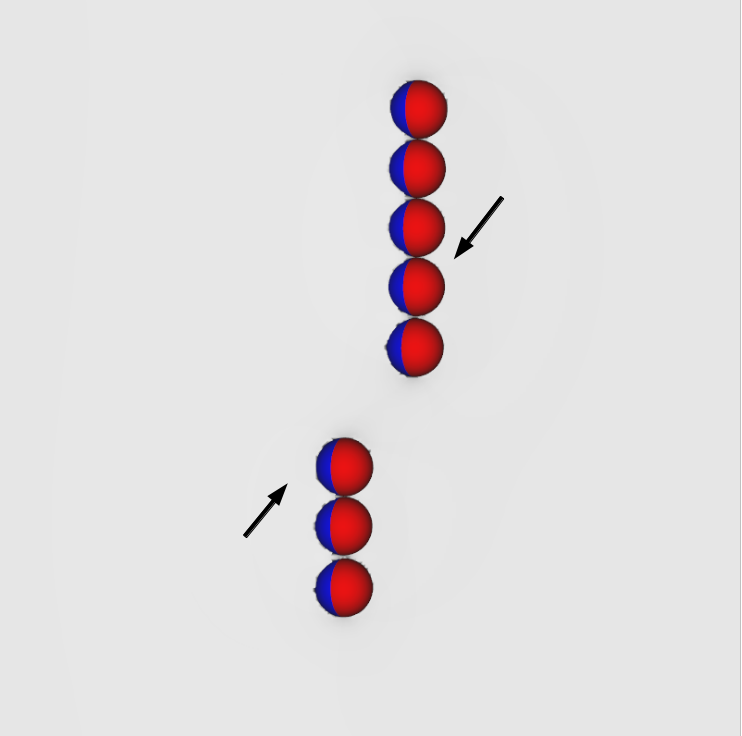}
 \subcaption{}
 \label{fig:n8-3} 
 &\includegraphics[width= 0.18\textwidth]{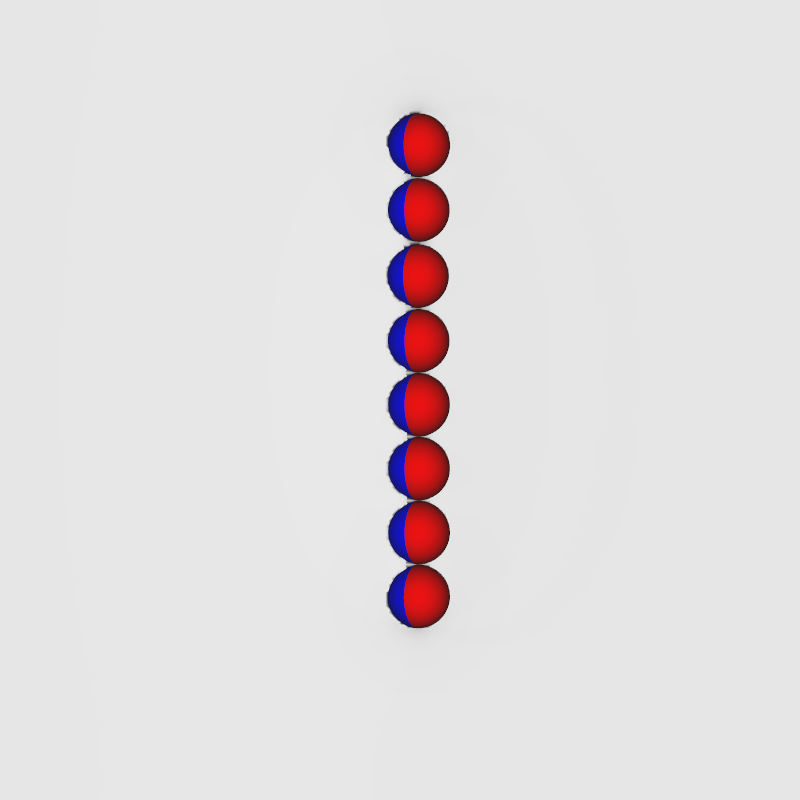}
 \subcaption{}
 \label{fig:n8-4} 
\end{tabular}
\end{center}
\caption{Snapshots of the assembly process of $8$ Janus particles adsorbed at a
fluid-fluid interface. (a) The particles are initially placed randomly distributed at the
interface. (b) The parallel external field is switched on, causing capillary
interactions between the particles, leading them to assemble into two separate
chains. Particles arrange side-by-side with bond angles $\varphi = 90^{\circ}$
within a chain. (c), (d) The two chains move relative to one another in the
direction shown by the arrows. (e) The separate chains merge into one chain
such that each particle has a bond angle $\varphi=90^{\circ}$ with its
neighbouring particles. 
}
\label{fig:n8}
\end{figure*}

\revisedtext{In order to understand the self-assembled structures of many-particles, it is required to consider the minimum energy orientation between two particles for a given tilt angle and separation.}
In the current case of equal bond angles $\varphi_A =\varphi_B= \varphi$,
minimising the total interaction energy with respect to the bond angle
using our theoretical model Eq.$\,$(A10) indicates that the interaction energy
decreases as the bond angle increases from $\varphi=0^{\circ}$ to
$\varphi=90^{\circ}$ (as shown in Fig.$\,8$ in Appendix A). 
This theoretical
analysis predicts that bond angles $\varphi_A=\varphi_B=90^{\circ}$ minimise
the interaction energy, and that there is no energy barrier stopping the
particles arranging into this configuration.

In order to test the predictions of our model, we performed simulations of two particles of radius $R=10$ separated by a distance $L_{AB}=40$ along the x-axis with total system size $S = 512 \times 96 \times 512$. We fix the tilt angles $\phi=90^{\circ}$ and measured the lateral force on the particles as the bond angle varies.
\figref{f-bond} shows that the capillary force is repulsive for bond angles $\varphi<50^{\circ}$ 
and attractive for bond angles $\varphi>50^{\circ}$. There is maximal attractive force between two particles for 
$\varphi_A =\varphi_B=90^{\circ}$. The simulation results show that for these parameters, two Janus particles with equal bond angles $\varphi_A = \varphi_B = \varphi=90^{\circ}$ minimises the interaction energy, agreeing with our theoretical predictions.
Moreover, the lateral force decreases monotonically with increasing bond angle indicating that there is no energy barrier stopping the particles achieving the minimum energy state. Therefore, two Janus particles of the kind investigated in this paper interacting as capillary dipoles should rearrange into a configuration with $\varphi=90^{\circ}$ bond angle.

\subsection{Multiple particles}

\begin{figure*}
\begin{center}
\begin{tabular}{m{0.15cm}m{3.4cm}m{3.4cm}m{3.4cm}m{3.4cm}m{3.4cm}}
 \rotatebox{90}{\hspace{1.5cm} $z$ $\longrightarrow$ }
&\includegraphics[width= 0.2\textwidth]{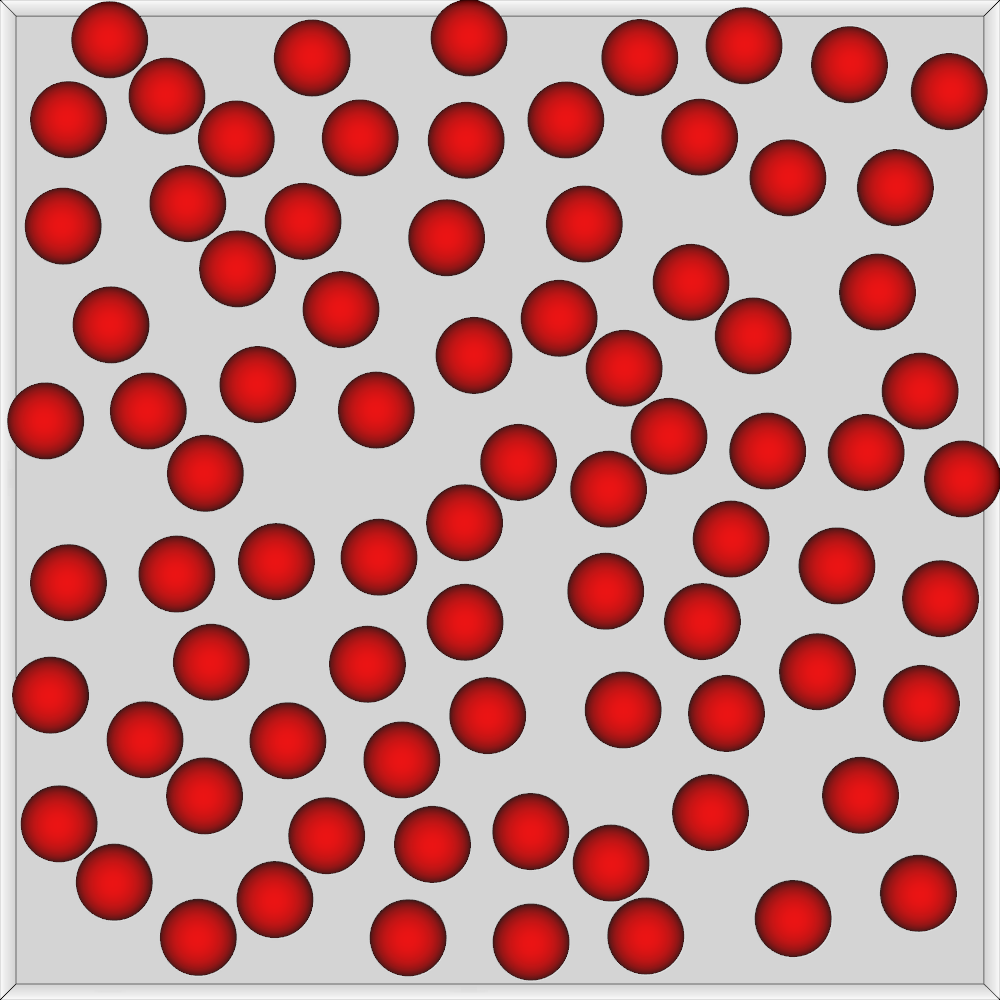}
\subcaption{$\bar{B}=0.0, Q=0.01$}
\label{fig:same_80_mt0} 
&\includegraphics[width= 0.2\textwidth]{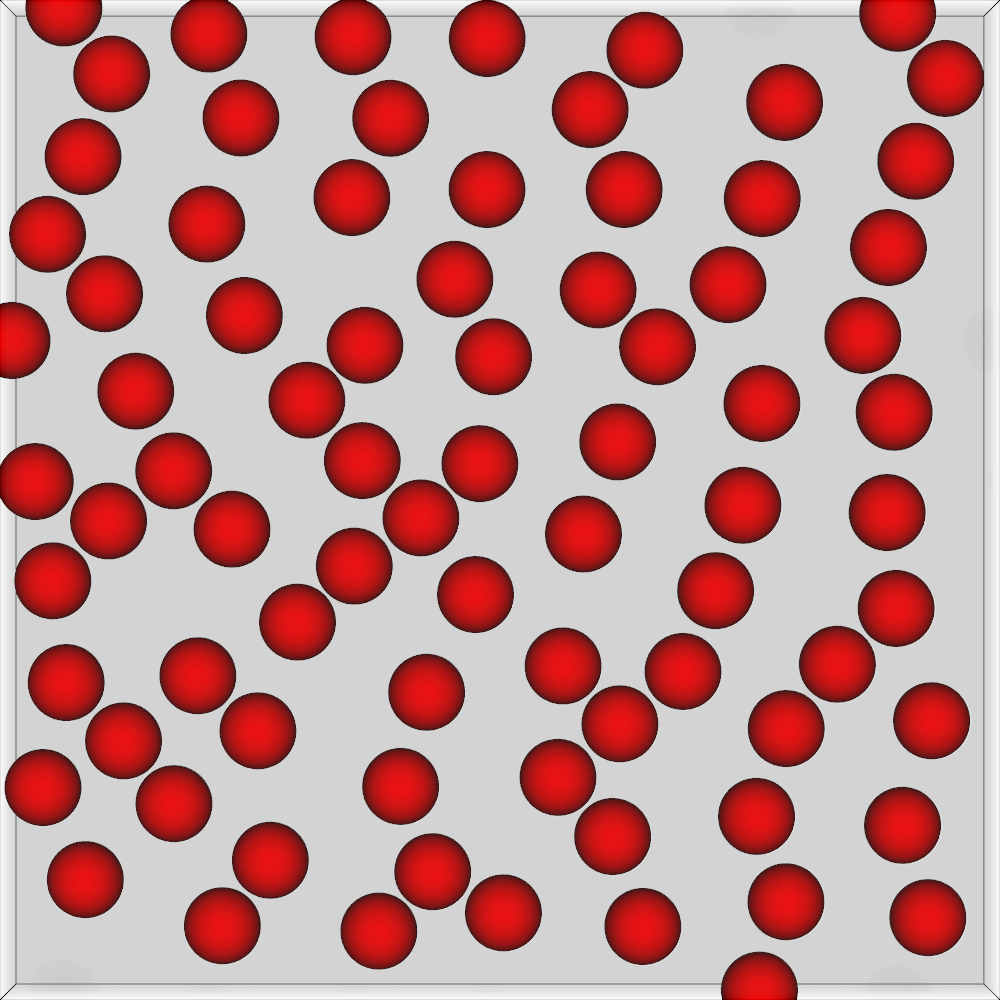}
\subcaption{$\bar{B}=0.13, Q=0.50$}
\label{fig:same_80_mt2} 
&\includegraphics[width= 0.2\textwidth]{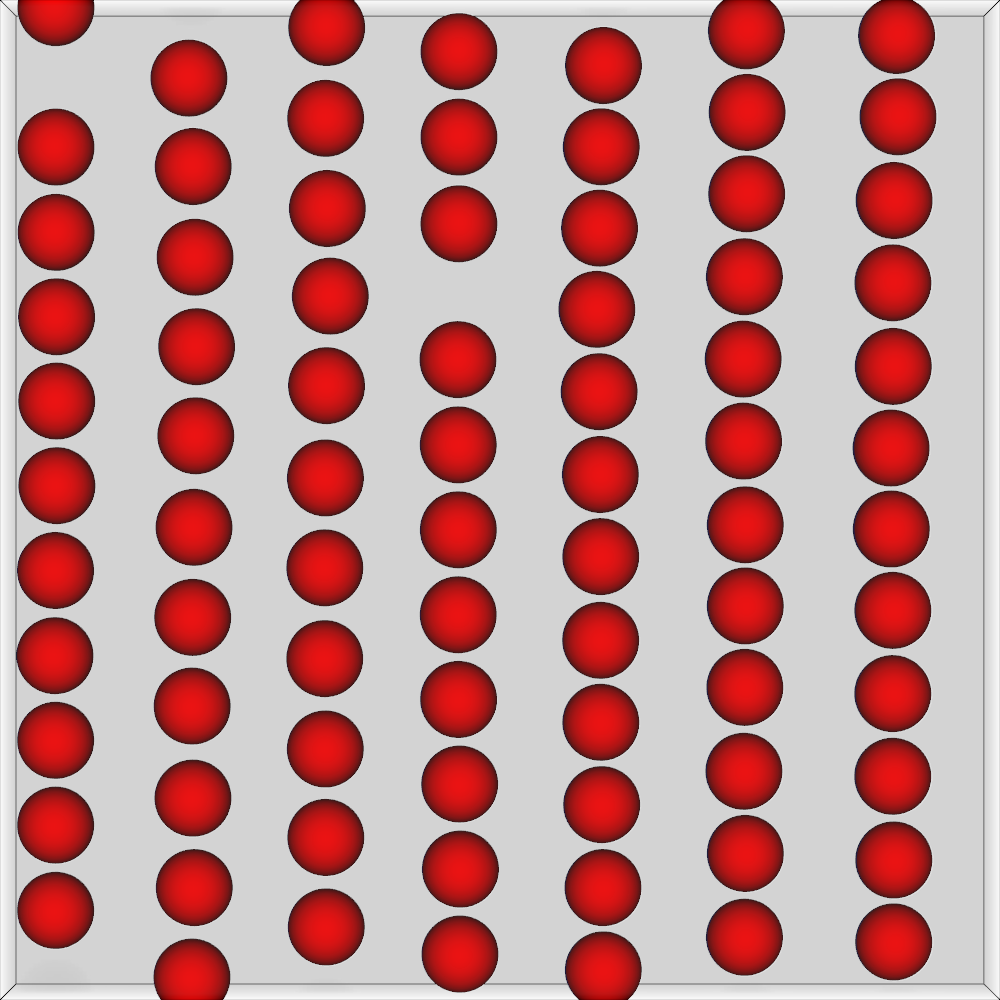}
\subcaption{$\bar{B}=0.30, Q=1.50$}
\label{fig:same_80_mt5} 
&\includegraphics[width= 0.2\textwidth]{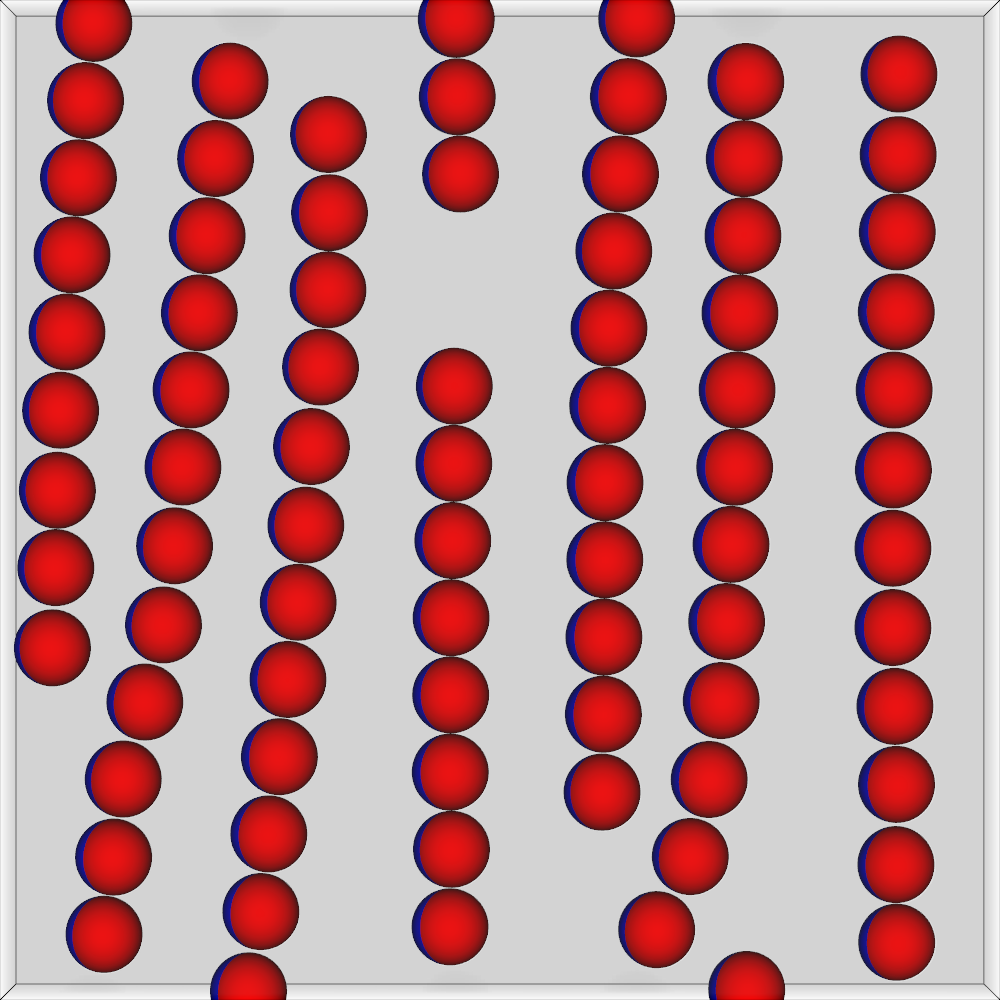}
\subcaption{$\bar{B}=0.65, Q=1.47$}
\label{fig:same_80_mt10} 
&\includegraphics[width= 0.2\textwidth]{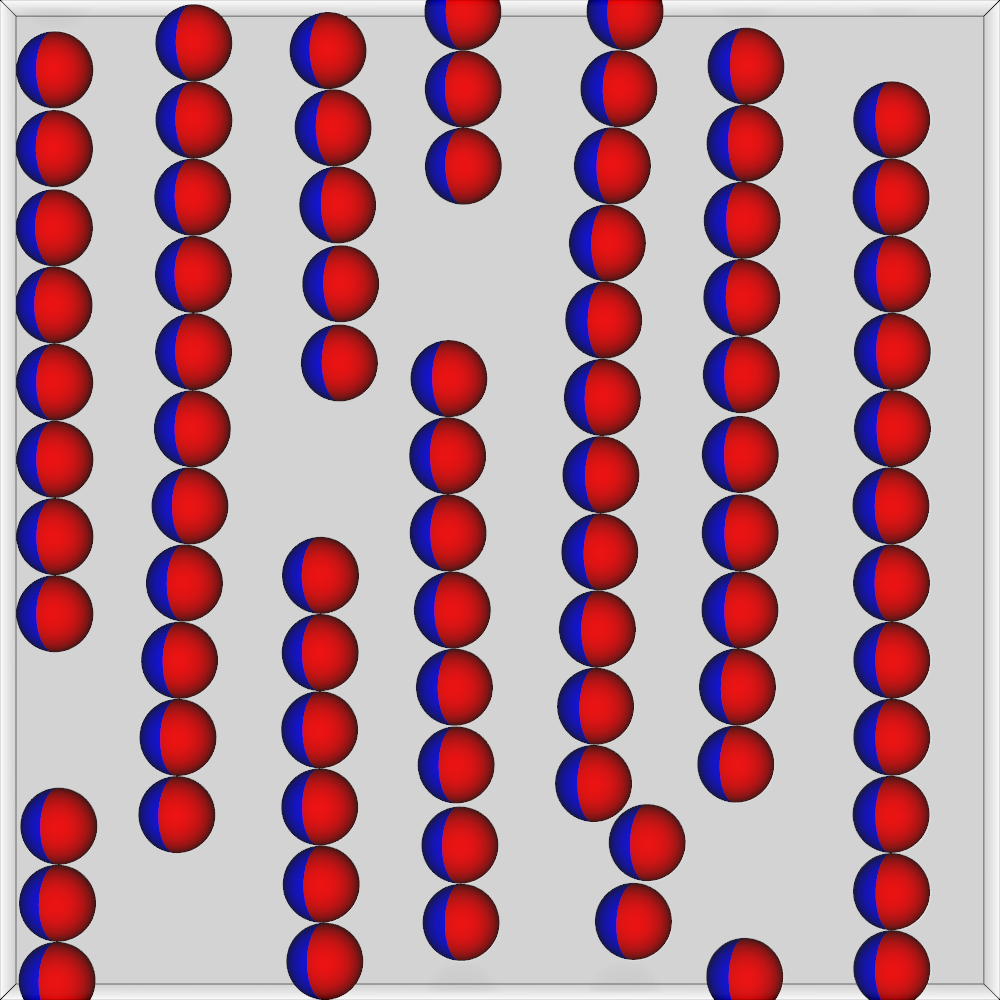}
\subcaption{$\bar{B}=1.31, Q=1.48$}
\label{fig:same_80_mt20} \\ [-0.9cm]

\end{tabular}
\end{center}
\caption{
Snapshots of self-assembled structures of a system with constant
surface fraction $\Phi= 0.38$ but varying dipole-field strength $\bar{B} =
B_m/\pi R^2\gamma_{12}$. (a) No external field applied. The particles distribute
randomly on the interface. (b) A small applied external field $\bar{B}=0.13$
shows the beginnings of chain formation but little global order. (c) Increase
the dipole-field strength to $\bar{B}=0.30$ creates a highly ordered lattice
structure.  (d) As the field strength increases to $\bar{B}=0.65$, the
interparticle distance within a chain decreases and the chains act as a
composite unit. Competing interactions with other chains causes chain bending
and curving. (e) For $\bar{B}=1.31$ the chains exhibit less curvature due to
the introduction of more defects in the system. The order parameter $Q$ is computed using Eq.~\ref{eq:order}.}
\label{fig:self_b}
\end{figure*}
In this section we study the arrangement and many-body dynamics of multiple Janus particles adsorbed at a flat fluid-fluid interface. We start by simulating 8 Janus particles each of radius $R=10$ adsorbed at an interface of area $A=256^2$.

\figref{n8} shows simulation snapshots of the assembly process for this system. The particles start off randomly distributed with no external field applied~(\figref{n8-0}). Once the external field is turned on, the particles arrange into two separate chains~(\figref{n8-1}). Within each chain, the particles arrange side--side with bond angles $\varphi=90^{\circ}$, in agreement with our pair-interaction analysis. Between the chains, the particles in one chain arrange with particles in the other chain such that their bond angles are $\varphi = 0^{\circ}$. 

Once arranged into individual chains, the chains act as a composite unit and move relative to one another. First, they move in opposite directions, as indicated by the arrows in~\figref{n8-1}. Once there is no end-end overlap of the chains, they begin to move towards one another (\figref{n8-2} and~\figref{n8-3}) before finally assembling into a single chain in which all particles are arranged side--side with bond angles $\varphi=90^{\circ}$ [Movie S1]. We need to reiterate that this assembly process occurs purely due to capillary interactions.

These results agree with our pair interaction analysis and suggest that short-range many-body effects are perhaps less relevant than in other capillary-interaction systems,
in which many-body structures do not correspond with the predicted structures from pair-wise interactions alone.\revisedtext{~\cite{Gary2015,supp} }

To investigate the assembly process further, we increase the number of particles on the interface. We define a surface fraction $\Phi= \frac{\pi N R^2}{A}$, where $N$ is the total number of particles and $A$ is the 
interface area before particles are placed at the interface.

\figref{self_b} shows the structure of particle monolayers with surface fraction $\Phi = 0.38$ that form as we vary the dipole-field strength $B_m$. Initially, the particles are distributed randomly on the interface with no external field applied~(\figref{same_80_mt0}). We then apply an external magnetic field directed parallel to the interface that switches on capillary interactions, as described previously, and we then allow the system to reach a steady state. 

For a dipole-field strength of $\bar{B}=B_m/\pi R^2\gamma_{12}=0.13$~(\figref{same_80_mt2}), we see some
ordering, but no formation of distinct chains in which particles are aligned
side-side. As we increase the dipole-field strength to
$\bar{B}=0.30$~(\figref{same_80_mt5}), the particles arrange into well-defined
chains. For this dipole-field strength, the chains exhibit little bending or
curvature. Within the chains, particles arrange with bond angles $\varphi =
90^{\circ}$, and particles arrange with particles in other chains with bond
angle $\varphi=0^{\circ}$, similar to the arrangements that we observed with 8
particles in~\figref{n8}. \revisedtext{Due to the periodic boundary conditions of our simulations, 
we observe coexistence of hexagonal and rectangular particular arrangements with neighbouring particles in other chains.} 
One can clearly observe a high degree of order for
this field strength. We also notice that the particles within the chains
maintain a clear separation between one another. Increasing the field strength
further to $\bar{B}=0.65$ ~(\figref{same_80_mt10}) and
$\bar{B}=1.31$~(\figref{same_80_mt20}) reduces the inter-particle distance between
particles within a particular chain, and the chains show a larger degree of
bending and curvature.

We suggest that the reason we observe a highly ordered lattice structure for
dipole-field strength $\bar{B}=0.30$~(\figref{same_80_mt5}) is due to the
strength of capillary interactions for this dipole-field strength; particles
are able to leave their chain and join other chains easily if it is
energetically favourable [Movie S2]. In contrast, for higher dipole-field
strengths $\bar{B}=0.65$ ~(\figref{same_80_mt10}) the capillary interactions between
the particles are stronger, as evidenced by their smaller inter-particle
separations within chains. Particles are more strongly bound to their initial
chains, and chains essentially become self-contained. These chains interact
with other chains as a composite unit, leading to the bending and the curving
of the chains, but particles are not easily able to leave one chain and attach
to another [Movie S3]. As the dipole-field strength increases yet further to
$\bar{B}=1.31$~(\figref{same_80_mt20}), the chains exhibit less bending and curvature
due to the introduction of more ``defects'' in the structure. Those appear due to the fact that
there is more available interface area because of the smaller inter-particle
separations. 

Our results suggest that for intermediate dipole-field strengths there is a ``sweet-spot'' capillary interaction magnitude that allows the rearrangement of the particles into energetically favourable structures. For field strengths of this magnitude, it may be possible to create thermodynamically stable monolayers, as opposed to the meta-stable monolayers one usually observes in monolayers of particles interacting via capillary interactions~\cite{Gunther2013}. 

\begin{figure}[b]
\includegraphics[width= 0.5\textwidth]{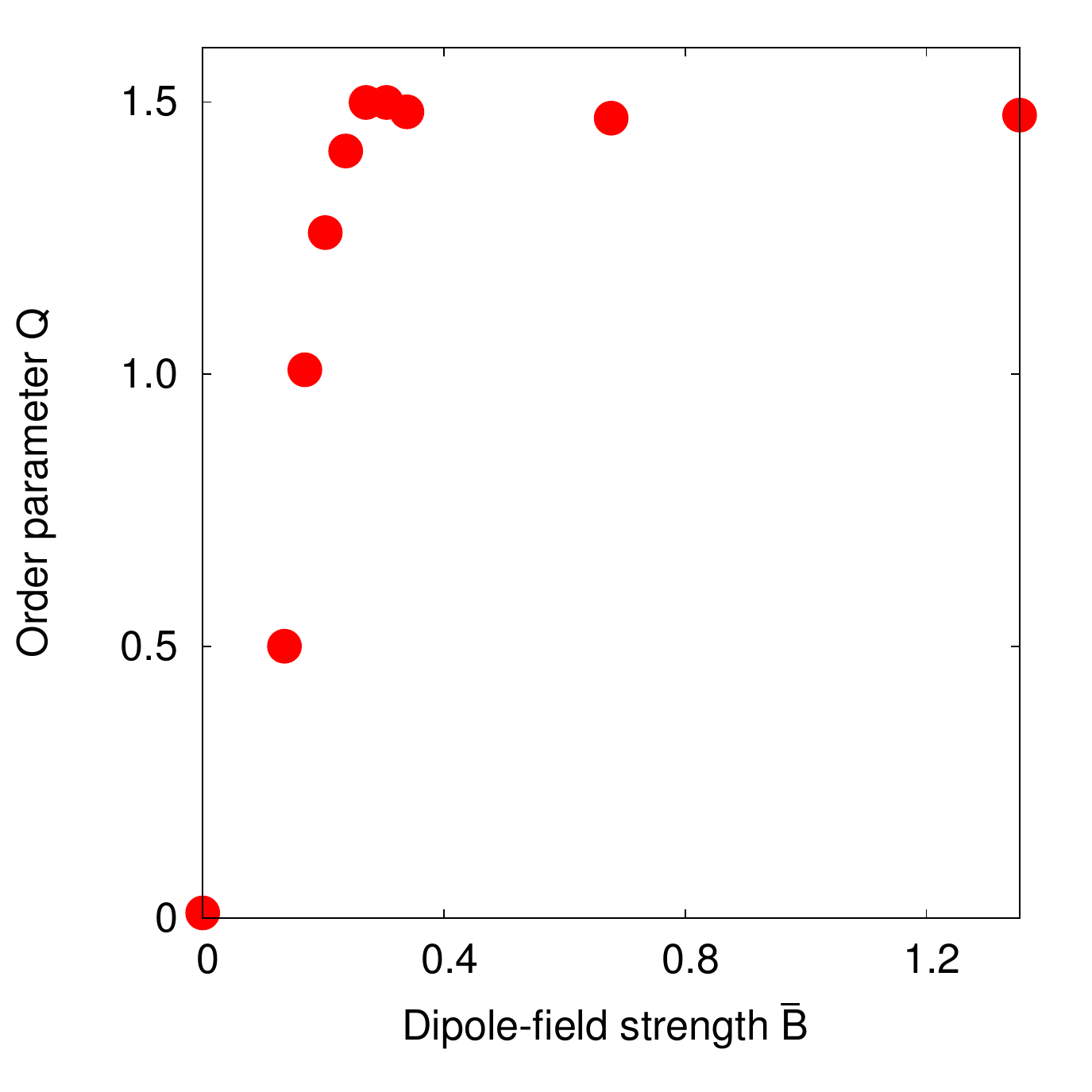}
\caption{Order parameter $Q$ for a surface fraction $\Phi= 0.38$ as the
dipole-field strength $\bar{B} = B_m/\pi R^2\gamma_{12}$ varies. The order
parameter $Q$ increases to $1.5$ for dipole-field strength $\bar{B} = 0.30$,
indicating that the particles form straight chains.
}
\label{fig:order}
\end{figure}

In order to characterize the straightness and order of the chains, we introduce a pair-orientation order parameter $Q$, defined as
 \begin{equation}
  Q = \frac{\sum_{k=1}^N \sum_{j=1}^{M_k} \frac{3\cos(2\psi_k^j)}{2}}{\sum_{k=1}^N M_k}
\label{eq:order}
 \end{equation}
where $N$ is the number of chains, $M_k$ is the number of particle pairs in chain $k$, and $\psi$ is the angle between the center-to-center vector of the particle pair $j$ and the $z$ axis. The centre-to-centre vector of the pair $j$ is the vector connecting centres of two adjacent particles $j$ and $j+1$.
The order parameter $Q$ takes values $1.5$ for chains whose pair vectors are parallel or $-1.5$ for chains whose pair vectors are perpendicular to the $z$ axis, respectively. 

\figref{order} shows the value of the order parameter $Q$ as the dipole-field strength $\bar{B}$ increases. We see that the order parameter increases monotonically before reaching its maximum value $Q=1.5$ at a dipole-field strength $\bar{B}\approx 0.3$. As the dipole-field strength increases further, the order parameter slightly decreases before reaching a constant value $Q \approx 1.47$ for dipole-field strengths $\bar{B} > 0.4$. For these order parameter values, the particles within chains should be parallel with one another, which agrees with our observations in~\figref{self_b}. Further, the maximum $Q$ value we observe in~\figref{order} at dipole-field strength $\bar{B}=0.30$ reconciles with our results in~\figref{self_b}.

\figref{order} suggests that the transition to a highly-ordered state is a smooth, second-order phase transition, rather than a first-order transition as observed with non-Janus ellipsoidal particles,~\cite{Gary2014a,Gary2014b} due to the absence of any energy barrier separating particular particle configurations.

\section{Conclusion}
\label{sec:conclusions}

We studied capillary interactions between magnetic spherical Janus particles both theoretically and numerically. Capillary interactions between magnetic spherical Janus particles are induced by applying a magnetic field parallel to the interface, which causes the particles to tilt and to deform the interface. We derived an analytical model for the interaction between two such particles using the superposition and small interface deformation approximations. Our model predicts that the strength of capillary interactions should rapidly increase for small particle tilt angles before reaching a constant value. It also predicts that a bond angle $\varphi = 90^{\circ}$ corresponding to the side--side configuration between two particles minimises the interaction energy between the particles, and that there is no energy barrier prohibiting the particles from achieving this configuration. 

We carried out lattice Boltzmann simulations of two Janus particles adsorbed at a fluid-fluid interface that confirmed our theoretical predictions. We then investigated the dynamics and steady-state behaviour of monolayers of Janus particles. Our simulations revealed interesting dynamical behaviour, namely that particles like to arrange in long, straight chains in the side--side configuration. However, if there are many particles on the interface, steric interactions lead to these chains bending and an increase in the number of defects in the monolayer. 

Interestingly, we find that for intermediate dipole-field strengths a highly-ordered crystal-like arrangement of Janus particles is possible. This is because, for intermediate dipole-field strengths, the capillary interactions are weak, allowing particles to leave one chain and join another easily if it is energetically favourable to do so. In contrast, for high dipole-field strengths, capillary interactions are strong and particles are tightly bound to the chain that they initially join, leading to meta-stable structures. 

Our results have implications for the directed self-assembly of particles adsorbed at fluid-fluid interfaces and the creation of ordered lattice structures of particle monolayers.

\begin{acknowledgments}
Q. Xie and J. Harting acknowledge financial support from NWO/STW (STW project
13291). We thank the High Performance Computing Center Stuttgart and the J\"ulich Supercomputing Centre for the allocation of computing time. 
\end{acknowledgments}
  \appendix 

\section{Theoretical analysis of pair interactions}
The interaction energy between two particles can be written as
\begin{equation}
 \Delta E = \Delta E_{ff} + \Delta E_{pf}  
 \mbox{,}
\end{equation}
where $\Delta E_{ff}$ is induced by the fluid-fluid interface and $\Delta E_{pf}$ is contributed by the particle-fluid interface.

Firstly, we consider the interaction energy contributed by the deformed
fluid-fluid interface.
$\Delta E_{ff}$ can be written as~\cite{Stamou2000} 
\begin{equation}
  \Delta E_{ff} = 2\int_{C_B} h_B (\mathbf{n} \cdot \nabla h_A) dC_B   
  \mbox{,}
\label{eq:ff}
\end{equation}
where $C_B$ is a closed curve at the meniscus of particle B, $\mathbf{n}$ is a
unit vector perpendicular to the boundary pointing away from the area of
integration and $h_B$ is the interface height profile~\cite{Xie2015},
\begin{equation}
 h_B (r_B,\vartheta_B) = \zeta_B \cos\left(\vartheta_B-\varphi_{B}\right) \frac{r_c}{r_B}
 \mbox{,}
\end{equation}
where $r_B,\vartheta_B$ are cylinder coordinates centered around particle B and $r_c$ is the radius where particle and fluid interact.

We choose a local coordinate system $(r_B,\vartheta_B)$ centered at particle B.
In this coordinate particle A is located at $(L_{AB},\pi)$.
Therefore, $h_A$ in this local coordinate is tranformed to
\begin{align} 
& h_A (r_B,\vartheta_B) = \frac{\zeta_A r_c\cos\left(\tan^{-1}\frac{r_B\sin\vartheta_B}{L_{AB}+r_B\cos\vartheta_B}-\varphi_{A}\right) }{\sqrt{(r_B\sin\vartheta_B)^2+(L_{AB}+r_B\cos\vartheta_B)^2}}  \mbox{.}\nonumber \\
 &
\end{align}
Assuming the distance $L_{AB}$ is larger than radius $r_c$, we do Taylor expansion of the field $\nabla h_A$ at $C_B$
\begin{align}
  &\nabla h_A (r_B,\vartheta_B) \approx \nabla h_A (\mathbf{0}) + \mathbf{r} \cdot (\nabla \otimes \nabla) h_A (\mathbf{0}) \nonumber \\ 
   &=  \zeta_A r_c L_{AB}^{-2} \begin{pmatrix} -\cos\varphi_A + 2r_c L_{AB}^{-1} \cos(\vartheta_B+ \varphi_A)  \\  \sin\varphi_B -2r_c L_{AB}^{-1}\sin(\vartheta_B+ \varphi_A) \end{pmatrix}
   \mbox{.}
\label{eq:taylor}
\end{align}

We insert~\eqnref{taylor} into~\eqnref{ff} and obtain
\begin{equation}
  \Delta E_{ff} =  2\pi \zeta_A \zeta_B r_c^2 L_{AB}^{-2} \cos (\varphi_A + \varphi_B)
  \mbox{.}
  \label{eq:eff}
\end{equation}
\revisedtext{\eqnref{eff} shows that the interface energy contributed by the fluid-fluid interface decays according to a $L_{AB}^{-2}$ law, which is 
consistent with previous studies~\cite{Danov2010, Gary2015}.}

Secondly we consider the interaction free energy contributed by the particle-fluid interface.
Following the superposition approximation approach, we can write $\Delta E_{pf}$ as
\begin{eqnarray}
 \Delta E_{pf}&=& -\gamma \sin\beta \bigg[ \int_{C_A} \left[ (h_B+h_A) - h_A \right] dC_A   \nonumber \\
 &&+   \int_{C_B} \left[ (h_A+h_B) -h_B \right] dC_B \bigg] \nonumber \\
 &=& - \gamma \sin\beta \left[ \int_{C_A}  h_B  dC_A +   \int_{C_B}  h_A  dC_B \right]   \mbox{.} \nonumber  \\
&& 
\label{eq:epf0}
\end{eqnarray}
We have
\begin{align} 
& \int_{C_B} h_A dC_B  =  \int_{C_B} \frac{\zeta_A r_c\cos\left(tan^{-1}\frac{r_B\sin\vartheta_B}{L_{AB}+r_B\cos\vartheta_B}-\varphi_{A}\right)}{\sqrt{(r_B\sin\vartheta_B)^2+(L_{AB}+r_B\cos\vartheta_B)^2}}  dC_B  \nonumber \\
&=  \int_{C_B} \frac{\zeta_A r_c r_B\sin\vartheta_B  \sin\varphi_A +\cos\varphi_A (L_{AB}+r_B\cos\vartheta_B) }{r_B^2+L_{AB}^2+2L_{AB}r_B\cos\vartheta_B} dC_B
\mbox{.}
 \end{align}
 Because of the anisotropic particle surface properties, the integral over $C_B$ is split into integrals on hydrophobic and hydrophilic hemispheres:
\begin{align}
& \int_{C_B} h_A dC_B  =  (\int_{\varphi_B - \pi/2}^{\varphi_B + \pi/2}  h_A r_c d\vartheta_B - \int_{\varphi_B + \pi/2}^{\varphi_B + 3\pi/2}  h_A r_c d\vartheta_B) \nonumber \\
  &= \frac{\zeta_A r_c^2 \sin\varphi_A }{L_{AB}} \log \frac{r_c^2+L_{AB}^2 + 2L_{AB}r_c\sin\varphi_B}{r_c^2+L_{AB}^2 - 2L_{AB}r_c\sin\varphi_B} \nonumber \\
  &- \frac{2\zeta_A r_c^2\cos\varphi_A} {L_{AB}} \tan^{-1}\frac{(L_{AB}+r_c)\cot\frac{\varphi_B + \pi/2}{2}}{L_{AB}-r_c} \nonumber \\ 
  &-\frac{2\zeta_A r_c^2\cos\varphi_A} {L_{AB}} \left( \tan^{-1}\frac{(L_{AB}+r_c)\tan\frac{\varphi_B + \pi/2}{2}}{L_{AB}-r_c} - \frac{\pi}{2}\right)  \mbox{.}
  \label{eq:epf1}
  \end{align}
\revisedtext{Interestingly, \eqnref{epf1} shows that the interaction energy contributed by the particle-fluid interface decays according to a $L_{AB}^{-1}$ law.}
By inserting~\eqnref{epf1} in~\eqnref{epf0}, we obtain $\Delta E_{pf}$.

\begin{figure}[htbp]
\includegraphics[width= 0.45\textwidth]{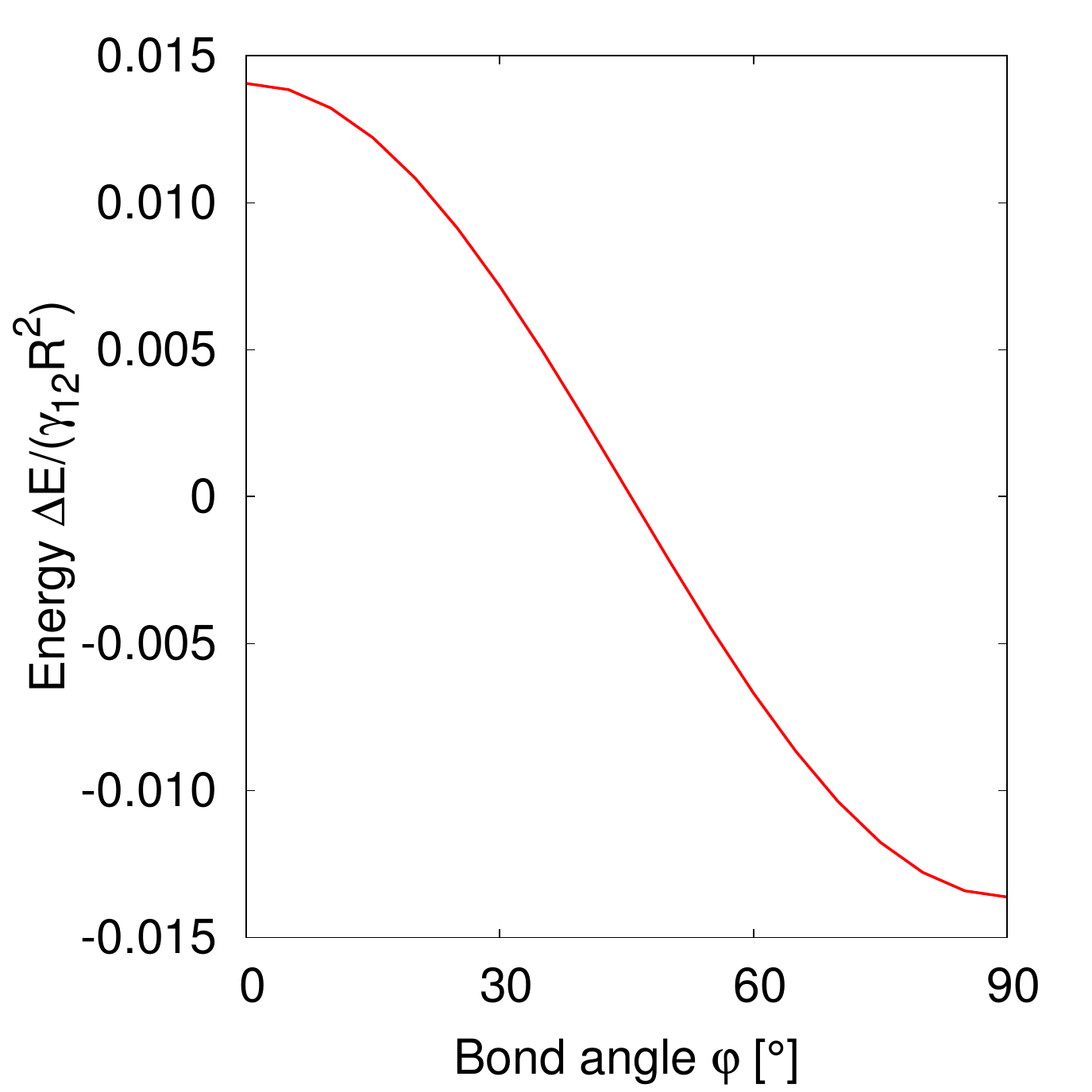}
\caption{Normalized interaction energy $\Delta E / (\gamma_{12}R^2)$ as a
function of bond angles
of particles with amphiphilicity $\beta = 21^{\circ}$. The two particles have a fixed tilt angle and a fixed pair sepration $L_{AB}/R=4$.
}
\label{fig:e-bond}
\end{figure}

Here we limit ourselves to discuss total energy for some special cases. 
By taking $r_c = R$, in the case $\varphi_A = \varphi_B = \varphi $ and $\zeta_A = \zeta_B =\zeta$, we obtain
\begin{align}
 &\Delta E  =  2\pi\gamma_{12} \zeta^2 R^2 L_{AB}^{-2} \cos (2\varphi) \nonumber \\
     &- \frac{ 2\gamma_{12} \sin\beta \zeta R^2 \sin\varphi }{L_{AB}}  \log \frac{R^2+L_{AB}^2 + 2L_{AB}R\sin\varphi}{R^2+L_{AB}^2 - 2L_{AB}R\sin\varphi} \nonumber \\
  &+\frac{ 4\gamma_{12} \sin\beta \zeta R^2\cos\varphi }{L_{AB}}    \bigg(\tan^{-1}\frac{(L_{AB}+R)\cot\frac{\varphi + \pi/2}{2}}{L_{AB}-R} \nonumber \\
   &+ \tan^{-1}\frac{(L_{AB}+R)\tan\frac{\varphi + \pi/2}{2}}{L_{AB}-R}- \frac{\pi}{2}\bigg)  
  & \mbox{.} \\\nonumber
  \label{eq:epf2}
\end{align}
We study the interaction energy as a function of the bond angle $\varphi$ of two neighbouring particles at a given tilt angle and at a fixed pair separation.
\figref{e-bond} shows that the interaction energy decreases with increasing bond angle from $\varphi=0^{\circ}$
to $\varphi=90^{\circ}$, which indicates that the minimum interaction configuration corresponds to a bond angle of $\varphi=90^{\circ}$.

In the case $\varphi_A = \varphi_B = 0^o$ and $\zeta_A = \zeta_B =\zeta$, 
the total interaction energy is 
\begin{eqnarray}
  \Delta E&= &2 \pi\gamma_{12} \zeta^2 R^2 L_{AB}^{-2} \nonumber \\ 
  &&+ \frac{8\zeta R^2 \gamma_{12} \sin\beta} {L_{AB}} \left(\tan^{-1}\frac{L_{AB}+R}{L_{AB}-R}-\frac{\pi}{4}\right) \mbox{.} \nonumber  \\
  &&  
 \label{eq:e00}
 \end{eqnarray}
 The lateral capillary force $\Delta F$ is the derivative of interaction energy with respect to $L_{AB}$, 
\begin{eqnarray}
  &&\Delta F = -4 \pi \zeta^2 \gamma_{12} R^2 L_{AB}^{-3}  \nonumber \\ 
   &&-  \frac{8\zeta R^2 \gamma_{12} \sin\beta} {L_{AB}^2} \left( \frac{L_{AB}R}{R^2+L_{AB}^2} + \tan^{-1}\frac{L_{AB}+R}{L_{AB}-R} - \frac{\pi}{4}\right)  \mbox{.}\nonumber \\
   &&
  \label{eq:force-upup}
 \end{eqnarray}

\end{document}